\documentclass[twocolumn,showpacs,preprintnumbers,amsmath,amssymb]{revtex4}

\usepackage{graphicx}
\usepackage{dcolumn}
\usepackage{bm}

\begin{document}


\title{Analysis on the effect of technical fluctuations on laser lineshape}
\author{Hyun-Gue Hong, Wontaek Seo, Moonjoo Lee, Wonshik Choi, Jai-Hyung Lee, and Kyungwon An}
\email{kwan@phya.snu.ac.kr}
\address{School of Physics, Seoul National University, Seoul, 151-742, Korea}
\date{\today}

\begin{abstract}
{We analyze theoretically the effect of technical fluctuations on
laser lineshape in terms of statistics of amplitude and phase noise
and their respective bandwidths. While the phase noise tends to
broaden the linewidth as the magnitude of the noise increases, the
amplitude noise brings out an additional structure with its spectral
density reflecting the magnitude of the noise. The effect of
possible coupling between those two noises is also discussed.}
\end{abstract}

\maketitle

\section{introduction}
Lineshape is one of the fundamental properties which tells about how
monochromatic a laser is. It also provides a way to investigate the
light emission mechanisms for various types of
lasers~\cite{Scully1997}.
However the technical fluctuations inherent in a laser and its
environment prevent one from observing spectral structure determined
by intrinsic dynamics only. In fact, the intrinsic limit of
linewidth set by quantum noise~\cite{Schawlow1958} is much smaller
than we usually observe in laboratory. The sources of such noise is
usually thermal or mechanical and they appear as, for example,
cavity drift, fluctuation of population inversion and instability
pump field. We may classify the aspects of those fluctuations into
phase and amplitude noise with various bandwidth. We analyze the
effect of technical fluctuations on lineshape in terms of amplitude
and phase in case of sinusoidal modulation and Gaussian
distribution. Their respective bandwidth and possible coupling
between those two~\cite{Vahala1983} are also included.

\section{phase noise}
Let us write the time variation of the electric field, the spectrum
of which we want to measure, as
\begin{equation} 
E(t)=E_{0}(t)e^{i[\omega_0 t+\phi(t)]}\;,
\end{equation}
where $E_{0}(t)$ is a slowly varying envelope and $\phi(t)$ is a
randomly fluctuating phase.

Firstly consider the case in which
the amplitude $E_0(t)$ is constant $E_0$ and $\phi(t)$ undergoes a
random walk process which imposes Gaussian statistics on $\phi(t)$.
The effective frequency is
\begin{equation} \omega_{eff}(t)=\omega_0 +{\dot \phi(t)}
\end{equation}
which represents that the time derivative of $\phi(t)$ makes jitter
around the carrier frequency. According to Wiener-Khinchin theorem,
the spectral lineshape is given by Fourier transform of the
first-order correlation function
\begin{equation} 
g^{(1)}(\tau)= \frac{\langle E^*(t)E(t+\tau) \rangle}
 {{\langle E(t) \rangle}^2}\;,
 \end{equation}
where $\langle \;\;\rangle$ denotes a time average. Using Eq. (1),
\begin{equation} 
\langle E^*(t)E(t+\tau) \rangle={|E_0|}^2e^{i\omega_0 \tau}
\langle e^{i \int_{t}^{t+\tau}  \dot{\phi}(t') dt'} \rangle\;. 
\end{equation}
For a normally distributed random variable $x$ with its probability density function (PDF) \begin{equation}
P[x]=\frac{1}{\sqrt{2\pi}\sigma} e^{-(x-\bar{x})^2/2\sigma^2}\;,
\end{equation} 
where $\bar{x}$ is the mean of $x$ and $\sigma^2$ is the variance, the characteristic function $\langle e^{i\omega x} \rangle $ is calculated to be $e^{i\bar{x} \omega} e^{-\frac{1}{2}\sigma^2 \omega^2}$~\cite{Papoulis2002}.
Taking $x=\int_{t}^{t+\tau}  \dot{\phi}(t') dt'$ and $\omega=1 $,
\begin{equation} 
\langle E^*(t) E(t+\tau) \rangle ={|E_0|}^2e^{i\omega_0 \tau}e^{-\frac{1}{2}\sigma_x^2}\;,
\end{equation}
where $ \bar{x}=0$ and $\sigma^2_x$ is given by
\begin{equation} 
\sigma_x ^2=\langle \int_{t}^{t+\tau} dt'
\int_{t}^{t+\tau} dt''  \dot{\phi}(t')\dot{\phi}(t'')\rangle\;.
\end{equation} 
The calculation of $\sigma_x^2$ depends on the bandwidth of the frequency jittering.

It is readily calculated if the spectrum of
$\dot{\phi}(t)$ is flat, that is, $\langle
\dot{\phi}(t)\dot{\phi}(t+\tau) \rangle$ is proportional to
$\delta(\tau)$. The constant of proportionality is found from
\begin{eqnarray}
\langle \dot{\phi}^2(t) \rangle &=& \int \dot{\phi}^2(t)
P[\dot{\phi}(t)] d\dot{\phi}(t) \nonumber\\&=& \frac{1}{\sqrt{2\pi}
\Delta R} \frac{\sqrt{\pi}}{2}(2 \Delta
R^2)^{\frac{3}{2}}\nonumber\\&=& (\Delta R)^2\;,
\end{eqnarray}
where $\Delta R$ denotes the magnitude of jittering and the PDF of
$\dot{\phi}(t)$
\begin{equation}
P[\dot{\phi}]=\frac{1}{\sqrt{2\pi}\Delta R}
e^{-{\dot{\phi}(t)}^2/{2(\Delta R)^2}}
\end{equation} 
was used. Then 
\begin{eqnarray} 
\sigma_x ^2&=&\int_{0}^{\tau} dt'
\int_{0}^{\tau} dt''  \langle \dot{\phi}(t')\dot{\phi}(t'')\rangle
\nonumber\\
 &=&(\Delta R)^2\int_{0}^{\tau} dt'
\int_{0}^{\tau} dt'' \delta(t'-t'') \nonumber\\&=&\sqrt{2}(\Delta
R)^2 |\tau|\;,
\end{eqnarray}
where, in the first line, $t$ is replaced with zero regarding the
noise being stationary. Finally we came to obtain 
\begin{equation}
g^{(1)}(\tau)=e^{i\omega_0 \tau}e^{- \frac{1}{\sqrt{2}}(\Delta R)^2
|\tau|}
\end{equation} 
and its Fourier transform gives a Lorentzian line
shape with FWHM of $(\Delta R)^2$.

If the spectrum of $\dot{\phi}(t)$ has a finite bandwidth,
$\sigma^2_x$ depends on the correlation time $t_c$ of $\dot{\phi}(t)$. We
can model it by a square-shaped temporal correlation as in Fig. 1(a). In terms
of frequency this noise is the white noise filtered by a
sinc-function low-pass filter of width$\sim t^{-1}_c$. The integration in Eq. (10) then takes the value depending on $\tau$ (Fig 1.(b))
\begin{equation}
\frac{\sigma^2_x}{(\Delta R)^2} = \left\{ 
\begin{array}{ll}
\tau^2 & \textrm{  if $\tau<\sqrt{2}t_c$}\\
2\sqrt{2}t_c\tau-2t^2_c & \textrm{  if
$\tau>\sqrt{2}t_c$}
\end{array} 
\right.
\end{equation}
Thus $g^{(1)}(\tau)$ is Gaussian up to $\tau=\sqrt{2}t_c$ and
thereafter exponentially decaying function. Of course it is
continuous at $\tau=\sqrt{2}t_c$ having the common value
$e^{-(2\Delta R)^2 t^2_c}$. If $t_c$ is longer than $(\Delta
R)^{-1}$, $g^{(1)}(\tau)$ is practically Gaussian. We have Gaussian
lineshape with its linewidth $\sim \Delta R$ in that case. In the
opposite limit where $t_c$ goes to zero, the Lorentzian lineshape
due to the white noise is recovered.

\begin{figure}
\includegraphics[width=3.2in]{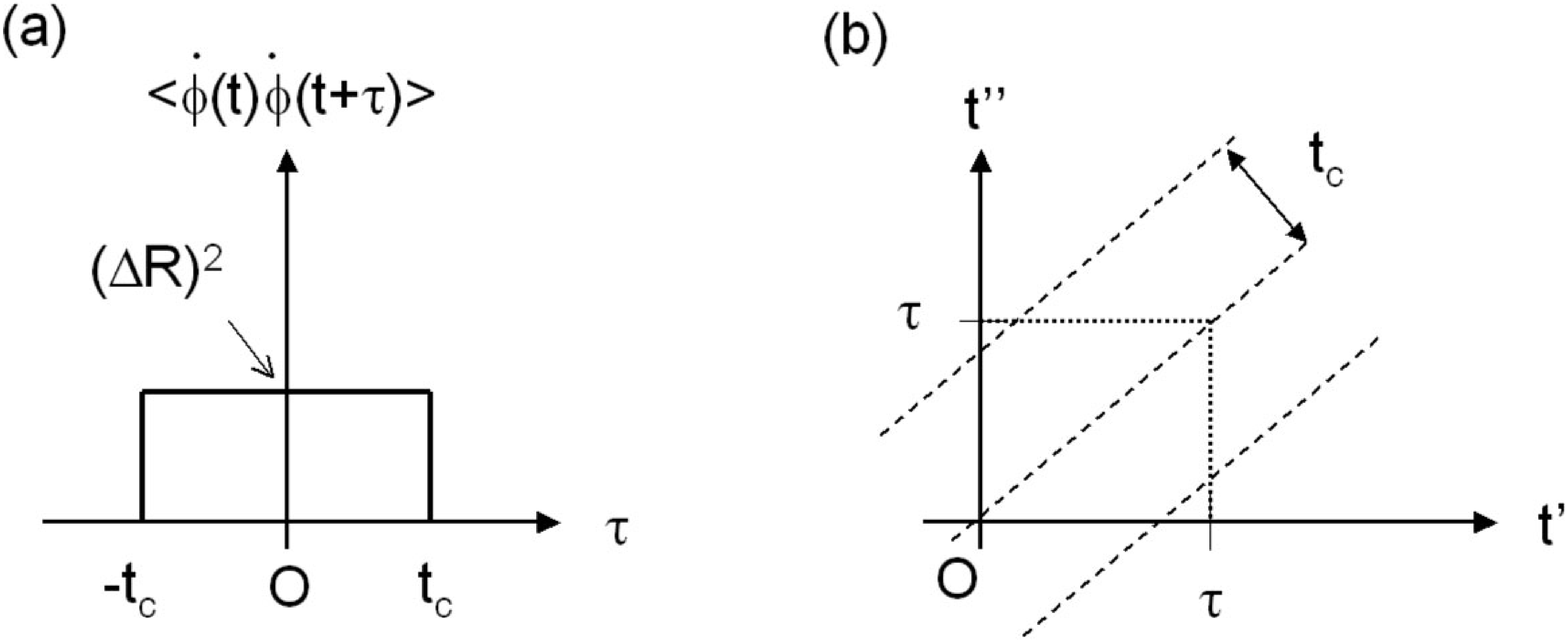}
\caption{(a) The correlation $\langle \dot{\phi}(t)\dot{\phi}(t+\tau)\rangle$ as a
function of $\tau$ as a model of finite bandwidth Gaussian noise.
(b) The value of integration of Eq. (10) is the area surrounded by
$t''=t'\pm \sqrt{2}$,$t'=\tau$ and $t''=\tau$ } 
\label{gaussian frequency}
\end{figure}

Another possible case of frequency jittering is slow modulation of
carrier frequency as it usually happens in a laser with its cavity
slowly drifting around the resonance frequency. PDF of Eq. (9) is no
more valid in such cases. Rather we start from
\begin{equation} 
\dot{\phi}(t)=\Delta F \cos(\Omega t)
\end{equation}
and accordingly 
\begin{equation} 
\phi(t)=\frac{\Delta
F}{\Omega}\sin\Omega t\equiv \phi_0 \sin \Omega t\;,
\end{equation}
where $\Delta F$ is the amplitude of modulation and $\Omega$ is the
slow frequency. For this simple harmonic oscillation, PDF of
$\phi(t)$ is given by
\begin{equation} 
P[\phi(t)]=\frac{1}{\pi}\frac{1}{\sqrt{\phi^2_0-\phi^2(t)}}\;.
\end{equation} 
The expectation value of $e^{ix}$ will be calculated
using this PDF where
\begin{eqnarray}
x&=&\int_{t}^{t+\tau}\dot{\phi}(t')dt'=\int_{t}^{t+\tau}\Delta F
\cos \Omega t' dt'\nonumber\\&=& \phi_0
[\sin{\Omega(t+\tau)}-\sin\Omega t] \nonumber\\&=& \phi_0(\sin\Omega
t \cos \Omega \tau
+\cos\Omega t\sin \Omega \tau - \sin\Omega t)\nonumber\\
&=&\phi\times(-2\cos^2{\frac{\Omega\tau}{2}})\pm\sin\Omega\tau\sqrt{\phi^2_0-\phi^2}\;.
\end{eqnarray}
By inspecting the PDF in Eq. (15), we can recognize that $\phi(t)$
spends most of its time near $\phi(t)\simeq \phi_0$ thereby neglect
the second term in Eq. (16). Then
\begin{eqnarray}
\langle e^{ix}\rangle&\simeq&\int_{-\phi_0}^{\phi_0} P[\phi]
e^{-2i\phi\cos^2{\frac{\Omega\tau}{2}}} d\phi\nonumber\\
&=&\int_{-\phi_0}^{\phi_0}
\frac{1}{\pi}\frac{e^{-2i\phi\cos^2{\frac{\Omega\tau}{2}}}}{\sqrt{\phi^2_0-\phi^2}}
 d\phi\nonumber\\&=&J_0\left(2\phi_0\cos^2{\frac{\Omega\tau}{2}}\right)\;,
\end{eqnarray}
where the integration in the last line involving $J_0$, Bessel
function of the first kind is performed in Ref.\ \cite{Arfken1995}.
Therefore
\begin{equation}
g^{(1)}(\tau)=e^{i\omega_0\tau}J_0\left(2\phi_0\cos^2{\frac{\Omega\tau}{2}}\right)\;.
\end{equation}

The graph of $J_0(2\phi_0\cos^2 {\Omega\tau}/{2})$ is given in
Fig. 2(a) and (c) for different value of magnitude of modulation
$\Delta F$. It is infinite pulse train with the repetition rate
$(\Omega/2\pi)^{-1}$. Its Fourier transform constitutes frequency
comb whose width of envelope is determined by the inverse time
duration of the pulse, $T$. (Fig.1 (b),(d)) T gets shorter as we
increase $\Delta F$ because the argument of Bessel function changes
by larger amount for the same change of $\tau$.  This is reasonable
result since the more harsh we swing the carrier frequency, the
wider the spectrum should be. $\Omega$ determines the repetition
rate, i.e. density (degree of fine-tooth) of the comb.

\begin{figure}
\includegraphics[width=3.2in]{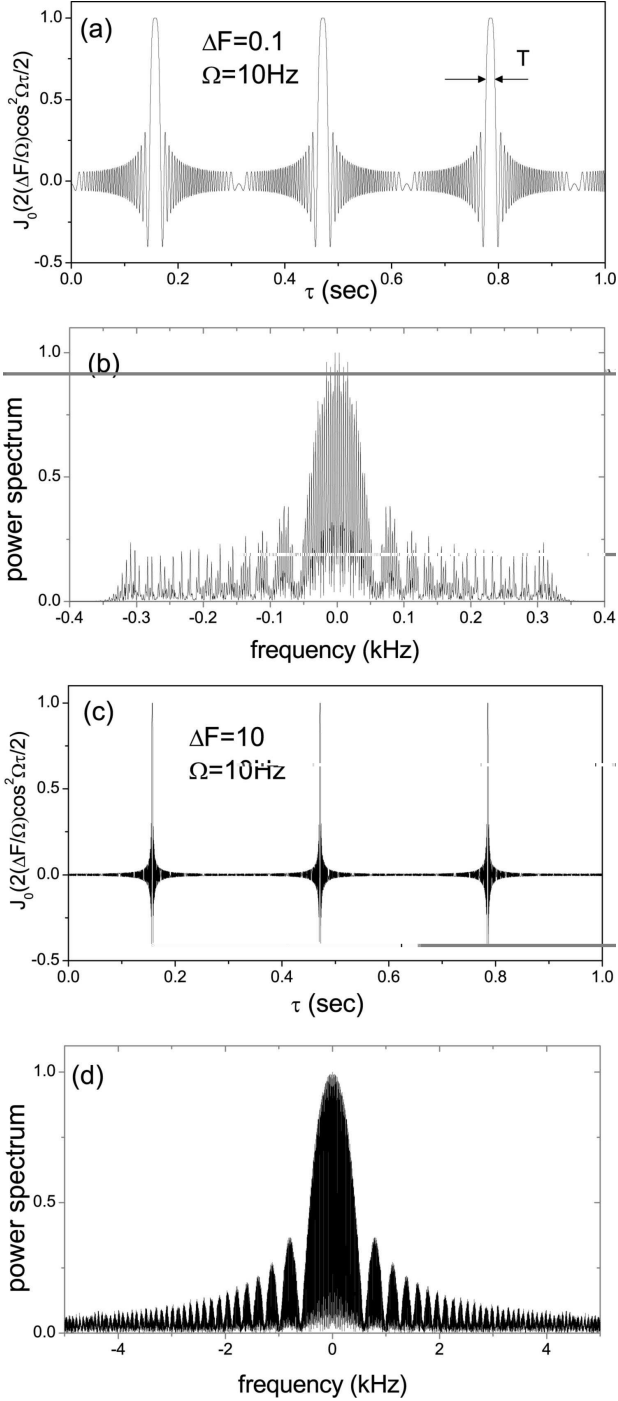}
\caption{(a) The factor $J_0(2\phi_0\cos^2{\frac{\Omega\tau}{2}})$ appearing in Eq.\  (18) as a function
of $\tau$. (b) Fourier transform of (a). (c) The factor $J_0(2\phi_0\cos^2{\frac{\Omega\tau}{2}})$ for $\Delta F$ 10
times larger than that in (a). (d) Fourier transform of (c).}
\label{sin mod}
\end{figure}

If another frequency components other than $\Omega$ is added in Eq.
(12), the fine spectral structure is easily destroyed by the
complexity of argument in Bessel function in Eq. (17). The
additional modulation is linearly added to $x$ and finally to the
argument of the Bessel function. This complicated argument brings
about reduction of the pulse height of $J_0$ by only a small number
of such superposition. Thus, in the spectrum, the spectral density
of each tooth in the comb gets smaller and interval between adjacent
teeth gets narrower.

\section{intensity noise}
Next let us include the effect of the amplitude fluctuation. The
criterion between amplitude and phase, in this analysis, rests with
their direct appearance in intensity. Consider the intensity
modulated like
\begin{equation}
{|E_{0}(t)|}^{2}=I_{0}(1+ M\cos\Omega t)\;,
\end{equation}
where the modulation depth $M$ is usually much less than 1. The
corresponding field amplitude can be written as
\begin{equation}
E_{0}(t)=\sqrt{I_0}(1+a_1e^{i\Omega t}+a_{-1}e^{-i\Omega
t}+a_{2}e^{i2\Omega t}\cdots\\)\;,
\end{equation}
where $1\gg a_{1},a_{-1}$ and $a_1,a_{-1}\gg a_2 \cdots\ $ for
$M\ll1$. This implies that the spectrum is given with sidebands,
which are symmetrically apart from the carrier frequency by integer
multiples of $\Omega.$ (Fig. 3(a)) Extending this idea to the noise
with finite band we can imagine superposition of many sidebands
comprised of all frequency component within the band. The resultant
spectrum contains the low-lying wing structure near the carrier
frequency. One example where the Gaussian noise is applied is
depicted in Fig. 3 (b). Note that the Lorentzian is smoothed out by
Gaussian profile which brings about the considerable deviation from
a Lorentzian lineshape.

\begin{figure}
\includegraphics[width=3.2in]{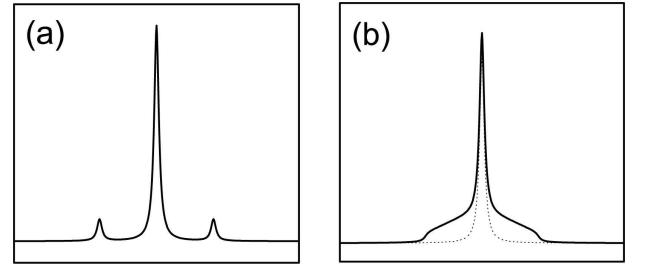}
\caption{(a) Sidebands due to the amplitude modulation. (b) If the
modulation has a finite band, the resultant spectrum has low-lying
wings corresponding to the noise band. In this figure, Gaussian
noise is presented.} 
\label{gaussian am}
\end{figure}

The explicit calculation starts with
\begin{equation} 
\langle E^*(t)E(t+\tau) \rangle=e^{i\omega_0 \tau}
\langle E_{0}^*(t)E_{0}(t+\tau)e^{i \int_{t}^{t+\tau}  \dot{\phi}(t') dt'} \rangle\;.
\end{equation}
Since
\begin{eqnarray}
E_{0}^*(t)E_{0}(t+\tau)&=&I_1(1+{|a_{1+}|}^2 e^{+i\Omega
\tau}+{|a_{1-}|}^2 e^{-i\Omega \tau}\nonumber\\& & +a_{1+}^* a_{2+}
e^{i\Omega (\tau-t)}+\cdots)\;,
\end{eqnarray}
each sideband make
$\bar{x}$ to shift by $0, \Omega, 2\Omega \cdots$ respectively while
$\sigma^2_x$ remains the same. Thus
\begin{eqnarray} 
g^{(1)}(\tau)&=&e^{i\omega_0 \tau}e^{-
\frac{1}{\sqrt{2}}(\Delta R)^2 \tau}\nonumber\\
& &\times\left(1+{|a_{1+}|}^2 e^{+i\Omega
\tau}+{|a_{1-}|}^2 e^{-i\Omega \tau}+\cdots\right)\;, \nonumber\\
\end{eqnarray}
so that 
\begin{eqnarray} 
{|{\cal E}(\omega)|}^2 &\sim&
\frac{1}{(\omega-\omega_0)^2+(\Delta R^2 /\sqrt{2})^2}\nonumber\\&&
+\frac{{|a_{1+}|}^2}{(\omega-\omega_0-\Omega)^2+(\Delta R^2
/\sqrt{2})^2}\cdots \;. \nonumber\\
\end{eqnarray}

The same result can be understood in a different way. The spectral
amplitude $\cal E(\omega)$ is the Fourier transform of the product
of two functions: $ E_{0}(t)$ and $e^{i[\omega t+\phi(t)]}$. Thus
the overall spectrum is given by convolution of each spectrum of the
functions. If the bandwidth of the amplitude noise is Gaussian,
\begin{equation} 
{\cal E}(\omega)\sim \int
 \frac{\delta(\omega' -\omega_0)
 +a_G e^{(\omega' - \omega_0)^2/{\Delta_G}^2}}{(\omega'-\omega-\omega_0)^2+\Delta_L^2}
\,d\omega'\;,
\end{equation}
where $\Delta_L$ and $\Delta_G$ are spectral widths of the
Lorentzian and Gaussian respectively and $a_G $ is relative spectral
amplitude of Gaussian noise.

If the bandwidth of amplitude fluctuation is too broad, the
low-lying structure may not be easily recognized. The manifestation
of amplitude noise can then be confirmed in the experiment from the
intensity correlation function $g^{(2)}(\tau)$ since
\begin{eqnarray}
\lefteqn{g^{(2)}(0)} \nonumber\\
&=&\langle(1+\int m(\Omega)\cos{\Omega t}\ d\Omega)(1+\int
m(\Omega ')\cos{\Omega 't} d\Omega
')\rangle\nonumber\\
&=&1+\frac{1}{2} \int d\Omega \int d\Omega m(\Omega)m(\Omega')\delta(\Omega-\Omega ')\nonumber\\
&=&1+\frac{1}{2} \int d\Omega m^2(\Omega)\;,
\end{eqnarray}
where $m(\omega)$ is modulation density. The peak near $\tau=0$
definitely reveal the effect of amplitude fluctuation.

There might be possible coupling between amplitude and phase as is well known in semiconductor lasers. If
$\phi(t)$ is affected by the intensity modulation as a quadrature
$\phi(t)=2\delta\sin{\Omega t}$ where $\delta$ designates the
magnitude of coupling,
\begin{eqnarray} 
E(t)&\simeq&(1+2a_1\cos{\Omega t})\cos{(\omega t
+2\delta \sin{\Omega t})}\nonumber\\&=&\frac{1}{2}e^{i\omega
t}+(a_1+\delta)e^{i(\omega+\Omega)}+(a_1-\delta)e^{-i(\omega-\Omega)}
\nonumber\\&&+c.c.
\end{eqnarray}
Hence this type of coupling brings about the asymmetry in the
sidebands.

\section{conclusion}

We investigated the effect of technical fluctuations on laser lineshape.
The Gaussian noise in frequency make the lineshape a Lorentzian or
a Gaussian depending on the correlation time of the noise. Slow swing
of frequency results in a lineshape with many sidebands by the
periodicity involved. The amplitude noise imposes the low-lying wing
structure in the spectrum. The bandwidth of the noise only
determines the width of the additional structure while the coupling
between the amplitude and phase might lead to an asymmetry in the structure.

\end{document}